%% file: main.tex
\documentclass[10pt,conference]{IEEEtran}

\usepackage{url}

\usepackage{fancybox}

\usepackage{cite}
\usepackage{booktabs}
\usepackage{tcolorbox}
\usepackage{xspace}
\usepackage{balance}

\usepackage{listings}

\usepackage{multicol}
 \usepackage{booktabs}
 \usepackage{multirow}
\usepackage{graphicx}

\ifCLASSINFOpdf
\else
\fi
\usepackage{enumitem}
\usepackage{balance}

\usepackage{algorithmic}

\usepackage{array}
\newcommand{\newtext}[1]{\textcolor{black}{#1}}

\newcommand{\etal}{et~al.\xspace}
\newcommand*{\RQone}{What are the data-access performance anti-patterns prevalent in data-intensive systems?}

\newcommand*{\RQtwo}{How do developers perceive the relevance of data-access performance anti-patterns?}

\begin{document}
%
\title{Data-access performance anti-patterns in data-intensive systems}



%

\author{\IEEEauthorblockN{Biruk Asmare Muse}
\IEEEauthorblockA{
Polytechnique Montréal\\
biruk-asmare.muse@polymtl.ca}
\and
\IEEEauthorblockN{Kawser Wazed Nafi}
\IEEEauthorblockA{Polytechnique Montréal\\
kawser.nafi@usask.ca}
\and
\IEEEauthorblockN{Foutse Khomh}
\IEEEauthorblockA{Polytechnique Montréal\\
foutse.khomh@polymtl.ca}
\and
\IEEEauthorblockN{Giuliano Antoniol}
\IEEEauthorblockA{Polytechnique Montréal\\
antoniol@ieee.org}}


\maketitle


\input{abstract}


%
\IEEEpeerreviewmaketitle

\section{Introduction}
\input{intro}
\section{Related work}
\input{related}

\section{Study method}
\input{studymethod}











%
\bibliographystyle{IEEEtran}
\bibliography{citation}

\end{document}

%% file: abstract.tex
\begin{abstract}


\textit{Context:} Data-intensive systems handle variable, high volume, and high-velocity data generated by human and digital devices. Like traditional software, data-intensive systems are prone to technical debts introduced to cope-up with the pressure of time and resource constraints on developers. Data-access is a critical component of data-intensive systems as it determines the overall performance and functionality of such systems. While data access technical debts are getting attention from the research community, technical debts affecting the performance, are not well investigated.

\textit{Objective:} Identify, categorize, and validate data access performance issues in the context of NoSQL-based and polyglot persistence data-intensive systems using qualitative study.

\textit{Method:} We collect issues from NoSQL-based and polyglot persistence open-source data-intensive systems and identify data access performance issues using inductive coding and build a taxonomy of the root causes. Then, we validate the perceived relevance of the newly identified performance issues using a developer survey.

Keywords- performance anti-patterns, data-intensive systems, polyglot persistence systems, NoSQL database, data-access code
\end{abstract}

%% file: intro.tex
Leveraging the vast amount of heterogeneous data generated by humans and machines to obtain actionable insights is becoming the center of attention in both industry\footnote{https://techjury.net/blog/big-data-statistics/\#gref} and the research community\cite{naeem2022trends}. Our decisions and way of life are greatly impacted by the insights and recommendations obtained from analyzing this big data \cite{Park2021}. Due to the size, heterogeneity, and complexity, handling this data with traditional software and applications is becoming more challenging. As a result, data-intensive software systems leveraging the availability of cloud infrastructures were introduced to address this challenge. The development of data-intensive systems involves the integration of data storage (e.g., Relational, NoSQL, and polyglot databases), processing (e.g., enrichment, classification, and prediction), and presentation (Services, APIs, and applications) frameworks. Data storage and access logic are critical components of data-intensive systems; determining their functionality, usability, and performance. Development of data-intensive systems posses several design, implementation, and quality assurance challenges \cite{CleveEtAl2010, Foidl2019,Hummel2018,Park2021}. Developers of such systems, like traditional software systems, could introduce technical debts \cite{tom2013exploration} that compromise software quality due to the release pressure and focus on the functional requirements. In addition to the technical debts common in traditional software systems, data-intensive systems could be prone to data access specific technical debts such as data-access smells (e.g., SQL code smell)\cite{muse2020prevalence} and data access performance anti-patterns \cite{shao2020database}.

While several studies investigated the prevalence and impact of technical debts in traditional software systems\cite{khomh2009exploratory,Fontana2013,Yamashita2013,johannes2019large,palomba2018diffuseness,iammarino2019self}, few studies considered data-intensive systems and technical debts related to data access code  \cite{de2019prevalence,muse2020prevalence,muse2022fixme,muse2022developers}. In particular, technical debts affecting the data access performance are not well investigated despite their importance in data-intensive systems. Recently, Shao \etal\cite{shao2020database} surveyed recent works to compile a list of data-access performance anti-patterns and complemented with new anti-patterns obtained from analyzing data access performance bugs from database-backed web applications. However, the existing performance anti-patterns are obtained by considering only relational or SQL databases and a small number of traditional web applications. Recently, NoSQL databases are becoming more popular components of data-intensive systems due to their horizontal scalability and handling of heterogeneous data (e.g., support for unstructured data and flexible schema) \cite{kleppmann2017designing}. Due to the differences in data access mechanisms between relational and NoSQL databases\cite{vathy2017uniform}, the performance anti-patterns identified in \cite{shao2020database} may not be generalized to NoSQL-based data-intensive systems. Furthermore, to leverage the advantages of both paradigms, polyglot persistence based systems are becoming more popular. While the advantage of choosing the database based on the nature of the data is clear, the associated design, implementation, and maintenance issues are not well investigated \cite{benats2021empirical}. Hence, a separate attention should be given to NoSQL-based and polyglot persistence data-intensive systems.

The goal of this empirical study is to specify data-access performance anti-patterns by conducting a qualitative analysis of issues collected from the repositories of 75 NoSQL-based and 87 polyglot persistence open-source data-intensive systems, and evaluating the perceived relevance of the performance anti-patterns by developers of industrial data-intensive systems. In particular, we seek to answer the following research questions.
\begin{enumerate}

    \item [RQ1] \textbf{\RQone}
    
    \textbf{\textit{Motivation:}} Specification of data access performance anti-patterns is important as it is the first step before building automatic detection tools that are leveraged to study the prevalence and impact of such anti-patterns on the quality of data-intensive systems and development of refactoring tools to help developers detect and fix the design issues before they impact the quality of their applications.
    \item [RQ2] \textbf{\RQtwo}
    
    \textbf{\textit{Motivation:}} Answering this research questions provides the perspective of the developers on the relevance of the newly identified performance anti-patterns. In addition, the taxonomy could be expanded with new potential anti-patterns that will be suggested by developers based on their experience. 
   
\end{enumerate}

The rest of the paper is organized as follows. Section \ref{sec:related} discusses the related work. Section \ref{sec:method} discusses in details the methodology that we plan to follow to answer our research questions. Finally, we outline our execution plan in Section \ref{sec:plan}. 

%% file: related.tex
\label{sec:related}
In this section, we discuss recent works related to data-access technical debts and performance anti-patterns.
\subsection{Data-access technical debts}
The specification \cite{karwin2010sql}, detection \cite{nagy2017static, Khumnin2017, Sharma2018}, prevalence \cite{muse2020prevalence,de2019prevalence, muse2022fixme} impact \cite{muse2020prevalence} and refactoring practices \cite{muse2022developers} of data-access technical debts have been investigated recently. Karwin \etal\cite{karwin2010sql} specified a catalog of SQL code smells. This work inspired researchers to investigate the smells and propose automatic detection approaches. The proposed tools perform a static analysis on SQL queries \cite{Khumnin2017, Sharma2018} and SQL queries embedded in data access code \cite{nagy2018sqlinspect} to detect SQL code smells. The availability of the tools promoted the empirical investigation of the prevalence and impact of data access technical debts in general and SQL code smells, in particular, using open source systems. De Almeida Filho \cite{de2019prevalence} investigated the prevalence and co-occurrence of SQL code smell on PL/SQL projects. In another work, Muse \etal\cite{muse2020prevalence} conducted a large-scale empirical study on SQL code smells using open-source data-intensive systems and found that SQL code smells are prevalent and persistent in data-intensive systems. In another work, Muse \etal\cite{muse2022fixme} investigated self-admitted technical debts (SATDS) in SQL and NoSQL-based data-intensive systems and found that data-access SATD is introduced as software gets more mature and persists for a long time without getting addressed. They also found that data-access SATDs are often introduced during bug fixing and refactoring activities and proposed a taxonomy of data-access SATDs. \newtext{
The authors identified a few instances of the self-admitted technical debts associated with query execution performance which can be considered as a data-access performance anti-pattern admitted by developers. In this study, however, we investigate data access performance issues using data from issues reported by the end users of NoSQL-based and polyglot persistence data-intensive systems.} Recently, developers' data-access code refactoring practices were also investigated recently using SQL-based data-intensive systems\cite{muse2022developers}. Most of the studies focus on relational or SQL-based systems. As a result, NoSQL-based and polyglot persistence systems did not get much attention. In this study, we focus on NoSQL-based and polyglot persistence based data-intensive systems to complement the state-of-the-art.

\subsection{Data-access performance anti-patterns}
There are few empirical studies on performance bugs \cite{jin2012understanding,nistor2013discovering,liu2014characterizing, yang2018not, shao2020database}, their root cause \cite{selakovic2016performance,jin2012understanding,shao2020database}, fixing strategy \cite{jin2012understanding,nistor2013discovering,shao2020database} their impact or relevance \cite{yang2018not,shao2020database} and both static and dynamic analysis based detection approaches \cite{nistor2013toddler,chen2014detecting,chen2016finding,yang2018powerstation}. Researchers also suggested various ways of data-access optimization to improve performance of database-backed web applications using caching and prefetching techniques. \cite{bowman2005optimization,cheung2012automatic,xiao2013context,chen2016cacheoptimizer}. Most of the performance studies either focus only on systems that use ORM driven relational databases or consider a subset of the performance anti-patterns, hence their findings may not be generalized to the case of NoSQL and polyglot persistence based data-intensive subject systems.

The closest work to this study is the work of Shao \etal\cite{shao2020database} where they provided a catalog of data-access performance anti-patterns obtained from (1) literature survey (24 anti-patterns) and (2) 10 new anti-patterns, analyzing real world performance bugs collected from seven open-source relational database backed web applications (BugZilla, DNN, Joomla!, MediaWiki, Moodle, WordPress, and Odoo). Our keyword based approach to identify data-access performance bugs is similar to the method used in this work. However, we extended the keywords to cover the case of NoSQL databases. While most anti-patterns are associated with SQL queries, some performance anti-patterns like moving computation to the sever or not caching could also be observed in NoSQL based systems. Hence, we aim to extend the data-access performance anti-patterns to the case of NoSQL and polyglot persistence data-intensive systems. In addition to providing the catalog, we will evaluate the perceived relevance of the data-access performance issues using a developer survey.

%% file: studymethod.tex
\begin{figure*}
\centering
\includegraphics[width=0.8\textwidth]{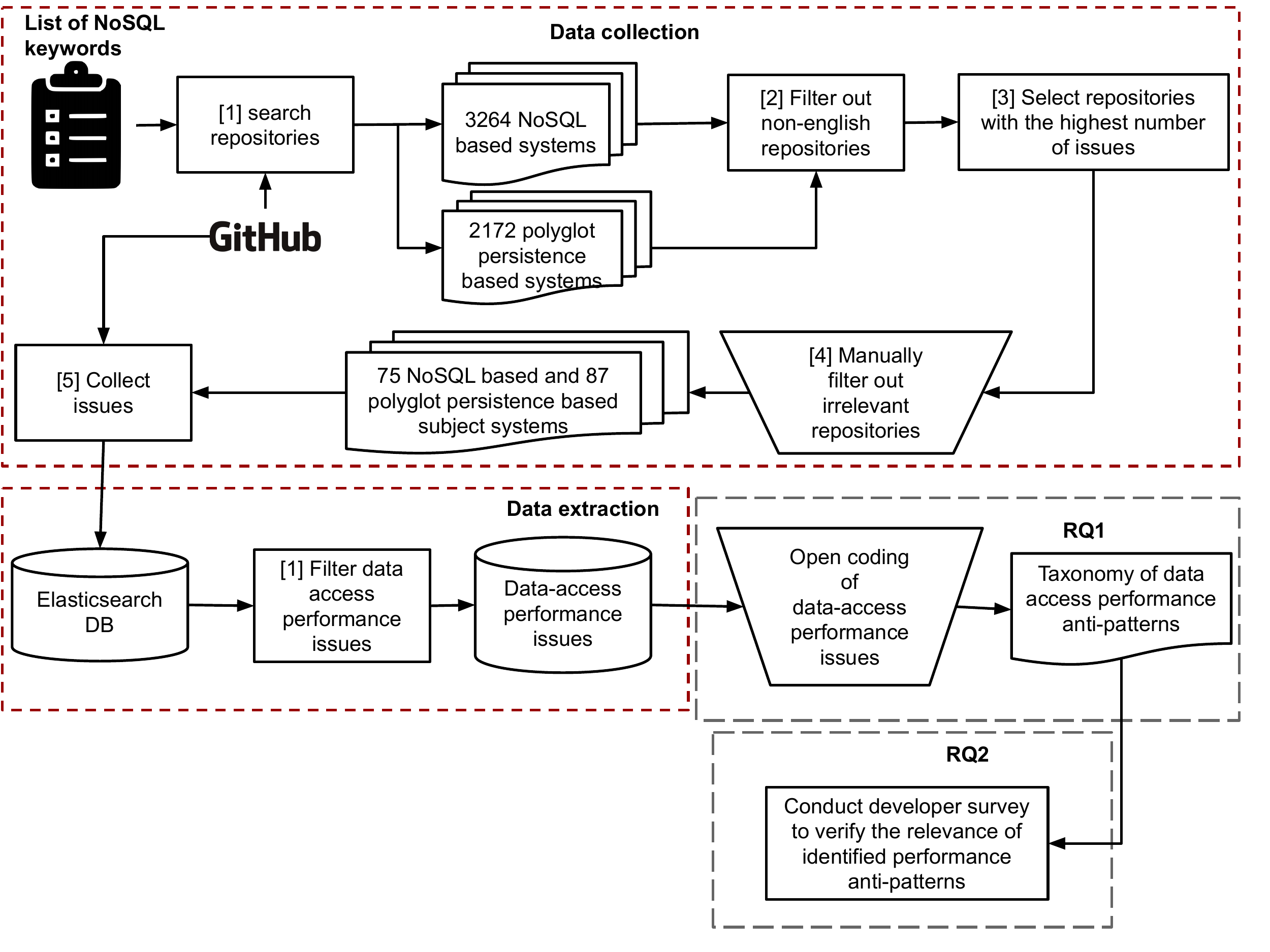}
\caption{Overview of the study method. }
\label{fig:Approach}
\end{figure*}

\label{sec:method}
The goal of this study is to identify, categorize and investigate the perceived relevance of data access performance issues in the context of NoSQL-based and polyglot persistence based data-intensive systems. In particular, we use issues collected from such subject systems as a data source for our analysis.  In this section, we describe the details of the study design including identifying subject systems, data collection, data analysis, and answering research questions. Figure \ref{fig:Approach} shows the overview of our study method.   

\subsection{Data collection}
We outline the steps we followed to select the 75 NoSQL and 87 polyglot persistence-based data-intensive subject systems in the following paragraphs.

\subsubsection{Search repositories}
 
Similar to the approach we followed in our previous work\cite{muse2022fixme}, we first collected a list of popular NoSQL databases such as MongoDB, Riak, and Redis. The list of databases is obtained from Eclipse JNoSQL framework\footnote{http://www.jnosql.org/docs/supported\_dbs.html} that provides a set of APIs for java applications to interact with NoSQL databases. The complete list of the databases is available in our replication package\cite{replication}. We conducted a repository search using GitHub rest API V3, we looked for repositories that mention at least one of the databases in our list in their title, description, or README file. To avoid tutorials and toy projects we set the search criteria to consider active repositories (whose latest push is not older than one year from the data collection date), repositories that are not mirrored to other repositories, repositories whose code size is at least 100 KB and repositories with at least two stars \cite{benats2021empirical}.

\begin{figure}
\centering
\includegraphics[width=\linewidth]{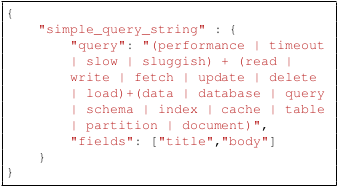}
\caption {Elasticsearch query to identify data access performance issues }
\label{lst:query}
\end{figure}

We ran the repository search on March 1, 2022, and we found 20340 candidate repositories before applying the quality filters and 3264 remained after applying the quality filter.
 
For identifying polyglot persistence data-intensive systems, we started with the work of Benats \etal \cite{benats2021empirical} where they investigated the usage of multi-database models using projects collected from Libraries.IO\footnote{https://libraries.io/}. They assessed the popularity of different database models including polyglot persistence models. As part of their replication package, they released the database of projects and their corresponding usage of relational and key-value, wide-column, document-oriented and graph-based NoSQL models. In particular, we used two tables from the provided database, the first one is a view (FILTERED\_REPOSITORIES\_VIEW) that provides the list of repositories after quality filtering (projects having at least 100 KB code size and at least two stars). This list contains 42176 projects. The second table (SQL\_NOSQL\_REPOSITORY\_WITH\_DBMS\_TYPE) provides the usage of the aforementioned database models in all projects. Combining the two, we obtained the database model usage of all the 42K systems. Since our goal is to identify systems that use polyglot persistence databases, we selected repositories that have at least two database models resulting in 6877 polyglot persistence-based applications. Once we collected the metadata of each repository, we removed repositories whose latest push is older than one year from the data collection date similar to the criteria of NoSQL subject systems. We finally ended up with 2172 polyglot subject systems.
 
 \subsubsection{Filter non-English repositories}
 We observed that the description and readme files are not written in English for some repositories. We filtered out repositories whose descriptions are not written in English using Langdetect\footnote{https://pypi.org/project/langdetect/} python library. After the language filtering, 2498 NoSQL and 1604 polyglot persistence-based repositories remained.
 
 \subsubsection{Select repositories with the highest number of issues}
 Since analyzing all issues from the 4102 repositories is not feasible and since there is no automatic way to measure the relevance of the systems, we need to restrict the number of subject systems by sorting the systems in decreasing order of the number of issues and picking the top projects. We selected the top 150 NoSQL based and 150 polyglot persistence-based systems as candidate subject systems.
 
 
 \subsubsection{Manually filter out irrelevant repositories}
 We manually went through each of the 300 subject systems' repositories on GitHub and investigated the description, code, README, and issues to understand their functionality and relevance to our study as data-intensive systems. We also filtered out repositories with non-English language descriptions missed by the Langdetect. After this filtering, the remaining 75 NoSQL based data-intensive systems and 87 polyglot persistence based data-intensive systems were selected as subject systems, 162 in total. The list of our subject systems is available in the replication package \cite{replication}.
 
 \subsubsection{Collect Issues}
 
 For each of the 162 subject systems, we collected all issues from their GitHub repositories using PyGithub\footnote{https://pygithub.readthedocs.io/en/latest/index.html}, which provides a python wrapper for GitHub REST API\footnote{https://docs.github.com/en/rest}. For each issue, we collected the title, body, issue number, URL, state, creation time, comments, labels, and closing time if the issue is closed. Since we want to investigate the solutions proposed to mitigate the issues, we are only interested in closed issues. We collected a total of 526672 closed issues from NoSQL based systems and 257840 closed issues from the polyglot persistence-based systems and stored them in the Elasticsearch database. Table \ref{tbl:subjects} provides the mean, standard deviation and the five number summary of the number of stars, number of forks, code size, and number of closed issues for our candidate subject systems.

\begin{table}[]
\centering
\caption{\newtext{Distribution of repository level metrics for NoSQL and Polyglot subject systems}}
\label{tbl:subjects}
\resizebox{\linewidth}{!}{%
\begin{tabular}{@{}lcccccccc@{}}
\toprule
\textbf{Metric} & \textbf{Group} & \textbf{Minimum} & \textbf{25\%} & \textbf{Median} & \textbf{Mean} & \textbf{75\% } & \textbf{Maximum} & \textbf{std } \\ \midrule
\multirow{2}{*}{Stars} & Polyglot & 5 & 71.5 & 508 & 4616.91 & 2896 & 61167 & 10747.29 \\
 & NoSQL & 7 & 405 & 2229 & 6083.81 & 9043.5 & 58918 & 9880.86 \\ \midrule
\multirow{2}{*}{Forks} & Polyglot & 0 & 29 & 127 & 1157.80 & 621 & 36393 & 4072.99 \\
 & NoSQL & 1 & 174 & 428 & 1318.23 & 954 & 21435 & 2863.24 \\ \midrule
\multirow{2}{*}{Code size (Kb)} & Polyglot & 744 & 7912 & 22,727 & 130741.72 & 138677.5 & 1719944 & 297229.78 \\
 & NoSQL & 1126 & 8515 & 23,010 & 132062.04 & 85,238.50 & 3836513 & 458862.46 \\ \midrule
\multirow{2}{*}{Closed issues} & Polyglot & 30 & 476 & 1028 & 2963.41 & 2163.5 & 33914 & 5541.40 \\
 & NoSQL & 303 & 1001 & 1936 & 6761.79 & 4366 & 128001 & 18274.97 \\ \cmidrule(l){1-9} 
\multirow{2}{*}{\newtext{Age (years)}} & Polyglot & 2.32 & 4.56 & 6.09 & 6.4 & 8.3 & 12.99 & 2.4 \\
 & NoSQL & 1.4 & 4.71 & 6.97 & 7.1 & 9.52 & 13.76 & 3.03 \\ \cmidrule(l){1-9} 
\end{tabular}
}
\end{table}

 \subsection{Data Extraction}
 In this subsection we discuss the procedure we followed to filter data access performance issues from the collected 526K issues and come up with the performance issue dataset.
 \subsubsection{Filter data access performance issues}
 
    Similar to the work of Shao \etal\cite{shao2020database}, we use the following heuristics to identify issues related to data access performance. We constructed an Elasticsearch query string using keywords associated with performance (performance, slow, timeout, sluggish), keywords associated with data access (read, write, fetch, update, delete, load), and database-related keywords (data, database, query, schema, index, cache, table, partition, document) connected by AND operator. The query is formed in multiple rounds by examining the returned results and modifying the query to minimize false positives. We applied the query against the issue title and the issue body. Figure \ref{lst:query} shows the final query we used to filter data access performance issues where $``|"$ represents OR operator and $``+"$ represents AND operator. We obtained 3760 issues from the NoSQL based systems and 2645 issues from the polyglot persistence-based systems.
    

 \subsection{RQ1 Taxonomy of data-access performance issues}

 We use open coding approach to come up with the taxonomy of data access performance issues. Two authors will participate in the labeling.  We leverage the relevance score obtained from the Elasticsearch to sort the performance issue dataset from most relevant to least relevant. The relevance score measures the relevance of the issues to the search query. We next perform the labeling of the sorted dataset in multiple rounds with 100 labels per round until label saturation is achieved. To minimize the impact of researcher bias, we assign the issues to each author under the constraint that each issue will be reviewed by two authors. Furthermore, each author will get the chance to label independently each issue and resolve labeling-conflicts by discussing among themselves. We utilize the list of performance anti-patterns \cite{shao2020database} as a seed and continue adding new labels as we continue the labeling.
 
 Once labeling saturation happens, all authors will participate in building the taxonomy from the ground up using the card sorting approach to come up with the taxonomy of data access performance issues. Leveraging the outcome of this research question, we will discuss the differences between the root causes of SQL based and NoSQl based data-access performance anti-patterns and the underlying reasons behind the difference. 

 \subsection{RQ2 Developer survey}
 
 Once we identify the data access performance issues taxonomy from  RQ1, the next step is to investigate the performance issues that are relevant to the developers of data-intensive systems. We plan to use a survey to collect the feedback of the developers on the proposed taxonomy of data access performance issues in NoSQL subject systems. The survey will consist of personal information queries about software development experience, preferred programming language, and developers' role in their organization or projects. This will help us to profile the professional experience and expertise of our respondents. We also introduce each category of the performance issues and ask the rating of their relevance from 1 to 5. We also put a place for their optional comments and an open question to mention new performance issue categories.

 \subsubsection{Survey participants selection}
 
 Our inclusion criteria are that developers should have at least one year of industrial development experience on NoSQL-based databases or polyglot persistence systems. We will use LinkedIn\footnote{https://www.linkedin.com/} as a platform to recruit survey participants. We first compile a list of keywords that will help us find NoSQL-based and polyglot data-intensive system developers. Then, we will use the keywords to search the profiles of the developers and manually go through the profiles of the matched participants if their experience is relevant to our study as per the inclusion criteria. We will also include contributors involved in the subject systems development as survey participants. \newtext{We plan to add a reward based on the random draw of willing participants that completed the survey. We expect at least 40 complete responses before closing the survey.}
 
 \newtext{Table \ref{tbl:survey_structure} shows the tentative survey questionnaire structure. The survey starts by introducing the research and asking for participants' informed consent. The following sections contain the anti-patterns associated with each high-level category and their description and a Likert scale of ratings of how critical the anti-pattern is. It ranges from: Not critical (1) to Highly critical (5). After that we will ask for optional additional comments about the smell. We plan to have one section per high-level category. Next, we will give a chance for participants to mention new types of performance anti-patterns with an optional open question. We end the survey by obtaining demographic information such as participants' experience in software development and data access code development.}

\begin{table}[]
\centering
\caption{Tentative survey structure template}
\label{tbl:survey_structure}
\resizebox{\columnwidth}{!}{%
\begin{tabular}{@{}lll@{}}
\toprule
\textbf{Section} &
  \textbf{Title} &
  \textbf{Content} \\ \midrule
1 &
  Introduction &
  \begin{tabular}[c]{@{}l@{}}- Summary of the description of the research, the role of participants, and the informed \\ consent to be accepted/denied by the survey\\ participants. The participants will access the remaining section on the condition that they \\ choose to participate in the research\end{tabular} \\
2 &
  Anti-pattern category \#1 &
  \begin{tabular}[c]{@{}l@{}}1. Anti-pattern \#1\\  - description \\  - rating (Not critical (1) to Highly critical (5))\\  - Comments(optional)\\ 2. Anti-pattern \#2 - description \\  - rating (Not critical (1) to Highly critical (5))\\  - Comments(optional)\\  ....\end{tabular} \\
\begin{tabular}[c]{@{}l@{}}3\\ ...\end{tabular} &
  \begin{tabular}[c]{@{}l@{}}Anti-pattern category \#2\\ ....\end{tabular} &
  \begin{tabular}[c]{@{}l@{}}1. Anti-pattern \#1\\  - description of the anti-pattern\\  - rating (Not critical (1) to Highly critical (5)) (Likert scale)\\  - Comments(optional)\\ 2. Anti-pattern \#2 - description \\  - rating (Not critical (1) to Highly critical (5))\\  - Comments(optional)\\  ....\end{tabular} \\
 &
  Additional anti-patterns &
  \begin{tabular}[c]{@{}l@{}}Would you like to suggest other type of data access performance anti-patterns that\\ are not mentioned in the previous sections? (Open question, optional)\end{tabular} \\
 &
  Demographics &
  \begin{tabular}[c]{@{}l@{}}How many years of experience do you have in software development?\\ How many years of experience do you have in back-end and \\ database access code development?\\ Which database access(persistence) frameworks, if any, \\ do you use for development?\\ Personal information(Email)(Optional)\end{tabular} \\ \bottomrule
\end{tabular}%
}
\end{table}

 \section{Execution plan}
 \label{sec:plan}
we outline our execution plan to answer the research questions in this work. Currently, we finished the data collection and data extraction stages of the study method. Hence, We will outline the steps we plan to answer the research questions.
\begin{enumerate}
    \item We prepare and test a web-based tool to help us in the labeling process.
    \item We start the first round of independent labeling using the labeling tool and conduct a discussion to resolve conflicts and reach a consensus on the labeling.
    \item We repeat the previous step in multiple rounds until we achieve label saturation.
    \item We organize a card sorting session to categorize and group the labels to build the taxonomy. We will be able to answer RQ1 after completing this step.
   
    \item Once we finish RQ1, we prepare the survey questionnaire and start to recruit the participants using a convenience sampling approach. 
    \item We will obtain research ethics approval to conduct 
    the survey from the relevant authority.
    \item We will share the survey with candidates that provided their consent to participate in the survey.
    \item Once we obtain the survey responses, we will perform quantitative and qualitative analyses of the candidate responses to answer RQ2.
\end{enumerate}

\section{\newtext{Limitations of the study}}

\newtext{One of the internal limitations of this study would be finding out and covering all the performance anti-patterns of data access systems. We are considering open source subject systems for our studies. Although we have selected data-intensive systems with the higher number of issues and carefully analyzed the available information with each of our subject systems, still there is a chance of missing some of the corner-case performance anti-patterns which were not been experienced by our selected subject systems. Moreover, at the time of studying with open-source subject systems, we can not ensure whether the findings of our study will be same for the industry-level data-intensive systems. To mitigate this, we plan to perform an extensive developer study. We plan to ensure at least 40 developers' participation for our study where the developers need to posses significant experiences with the data-access code and back-end development. And like our authors, developers will get a chance to add new performance anti-pattern labels based on their knowledge and experience.}

\newtext{In addition to that, there is a probability of missing some potential data-access performance issues as those were not appeared with our selected top 162 systems. Our subject systems are selected based on the criteria we described in the data collection, \textit{"selection of repositories with highest number of issues"}. To mitigate this limitation, we plan to give the opportunity to the developers who are going to participate in the survey to add non-listed data-access performance anti-pattern labels based on their own knowledge and experience.} 

\newtext{Another internal limitation could be ensuring the proper selection of issues from the subject systems for the study. It is evident that not all the issues are related to data-access performance anti-patterns. And some of the issues have improper descriptions too. To mitigate this challenge, we have decided to perform manual labeling of issues with all the authors' mutual decision and proper verification in several rounds. In addition to that, we will validate our identified and listed performance anti-patterns with developers.}

\newtext{The external limitations of our study could be the selection of experienced developers. To keep our study bias-free and ensure the quality of our study, we plan to select developers based on their deliberately declared experience with data access systems as well as their willingness to participate. As the first step of this process, we have decided to select developers for our study by sending invitations through LinkedIn.}